\documentclass[useAMS,usenatbib]{mn2e}

\usepackage{graphicx}
\usepackage{wasysym}
\usepackage{amssymb}
\usepackage{stmaryrd}

\title[WASP-14b: 7.3 M$_{\rm J}$ transiting planet in an eccentric orbit]
      {WASP-14b: 7.3 M$_{\rm J}$ transiting planet in an eccentric orbit}

\author[Y.~C.~Joshi et al.]{Y.~C.~Joshi$^{1}$\thanks{E-mail: y.joshi@qub.ac.uk},
D. Pollacco$^{1}$,
A. Collier Cameron$^{2}$, 
I. Skillen$^{3}$,
E. Simpson$^{1}$,
I. Steele$^{6}$,
\newauthor
R. A. Street$^{9}$,
H. C. Stempels$^{2}$,
D. J. Christian$^{1}$,
L. Hebb$^{2}$,
F. Bouchy$^{14}$,
N. P. Gibson$^{1}$,
\newauthor
G. H\'ebrard$^{14}$,
F. P. Keenan$^{1}$,
B. Loeillet$^{14,16}$,
J. Meaburn$^{10}$,
C. Moutou$^{16}$,
B. Smalley$^{5}$,
\newauthor
I. Todd$^{1}$,
R. G. West$^{7}$,
D. R. Anderson$^{5}$,
S. Bentley$^{5}$,
B. Enoch$^{4}$,
C. A. Haswell$^{4}$,
\newauthor
C. Hellier$^{5}$,
K. Horne$^{2}$,
J. Irwin$^{12}$,
T. A. Lister$^{12}$,
I. McDonald$^{5}$,
P. Maxted$^{5}$,
\newauthor
M. Mayor$^{13}$,
A. J. Norton$^{4}$,
N. Parley$^{4}$,
C. Perrier$^{15}$,
F. Pont$^{11}$,
D. Queloz$^{13}$,
\newauthor
R. Ryans$^{1}$,
A. M. S. Smith$^{2}$,
S. Udry$^{13}$,
P. J. Wheatley$^{8}$,
D. M. Wilson$^{5}$\\ \\ 
$^{1}$Astrophysics Research Centre, School of Mathematics \&\ Physics, Queen's University, University Road, Belfast, BT7 1NN, UK\\
$^{2}$School of Physics and Astronomy, University of St Andrews, North Haugh, St Andrews, Fife KY16 9SS, UK\\
$^{3}$Isaac Newton Group of Telescopes, Apartado de Correos 321, E-38700 Santa Cruz de la Palma, Tenerife, Spain\\
$^{4}$Department of Physics and Astronomy, The Open University, Milton Keynes, MK7 6AA, UK\\
$^{5}$Astrophysics Group, Keele University, Staffordshire, ST5 5BG, UK\\
$^{6}$Astrophysics Research Institute, Liverpool John Moores University, Twelve Quays House, Egerton Wharf, Birkenhead, CH41 1LD, UK\\
$^{7}$Department of Physics and Astronomy, University of Leicester, Leicester, LE1 7RH, UK\\
$^{8}$Department of Physics, University of Warwick, Coventry CV4 7AL, UK\\
$^{9}$Las Cumbres Observatory, 6740 Cortona Dr. Suite 102, Santa Barbara, CA 93117, USA\\
$^{10}$Dept of Physics and Astronomy, University of Manchester, UK\\
$^{11}$Physikalisches Institu, University of Bern, Sidlerstrass 5, 3012 Bern, Switzerland\\
$^{12}$Harvard-Smithsonian Center for Astrophysics, 60 Garden Street MS-16, Cambridge, MA 02138-1516, USA\\
$^{13}$Observatoire de Gen\`eve, Universit\'e de Gen\`eve, 51 Ch. des Maillettes, 1290 Sauverny, Switzerland\\
$^{14}$Institut d'Astrophysique de Paris, UMR 7095 CNRS, Universit\'e Pierre \&\ Marie Curie, 98$^{bis}$ bvd. Arago, 75014 Paris, France\\
$^{15}$Laboratoire d'AstrOphysique, Observatoire de Grenoble, Univ. J. Fourier - BP 53, F-38041 Grenoble Cedex 9, France\\
$^{16}$Laboratoire d'Astrophysique de Marseille, CNRS (UMR 6110), BP8, 13376 Marseille Cedex 12, France 
}
\begin{document}

\date{Accepted 2008 October 31. Received 2008 October 29; in original form 2008 September 10}

\pagerange{\pageref{firstpage}--\pageref{lastpage}} \pubyear{2008}

\maketitle

\label{firstpage}

\begin{abstract}
We report the discovery of a 7.3 M$_{\rm J}$ exoplanet WASP-14b, one of the most massive
transiting exoplanets observed to date. The planet orbits the tenth-magnitude F5V star
USNO-B1 11118-0262485 with a period of 2.243752 days and orbital eccentricity  $e = 0.09$.
A simultaneous fit of the transit light curve and radial velocity measurements yields a
planetary mass of 7.3$\pm$0.5 M$_{\rm J}$ and a radius of 1.28$\pm$0.08
R$_{\rm J}$. This leads to a mean density of about 4.6 g\,cm$^{-3}$ making it densest
transiting exoplanets yet found at an orbital period less than 3~days. We estimate this
system to be at a distance of $160\pm20$ pc. Spectral analysis of the host star reveals
a temperature of $6475\pm100$ K, log g = 4.07 cm\,s$^{-2}$ and $v\sin i = 4.9\pm1.0$
km\,s$^{-1}$, and also a high lithium abundance, $\log N({\rm Li})$ = 2.84$\pm$0.05.
The stellar density, effective temperature and rotation rate suggest an age for the
system of about 0.5--1.0 Gyr.
\end{abstract}


\section{INTRODUCTION}
The giant exoplanets that transit across the disks of their host stars are of great interest
due to their impact on our understanding of planetary structure.  Since the first discovery
of a transiting exoplanet HD209458b \citep{charbonneau, henry}, more than 40 transiting
systems have been found around nearby stars. Transit light curves, along with their radial
velocity motions, provide a wealth of information about the system including precise mass,
radius and mean density of the planet. This in turn allows us to probe their internal
structure by comparing their physical parameters with models of planetary structure and
evolution \citep{guillot, fortney}. Given the importance of these systems, several
wide-field surveys are in progress to find transiting exoplanets, e.g. HAT \citep{hat},
XO \citep{xo}, TrES \citep{tres}, and SuperWASP \citep{sw}. 

In this paper, we report the 14th exoplanet discovered in the SuperWASP survey which is
orbiting around a 10th-mag F5-type star in the Northern Hemisphere. The paper is organised
as follows. In \textsection 2, we give details of the discovery light curve and outcome of
our follow-up photometric and spectroscopic observations. Spectral analysis to determine
system parameters is described in \textsection 3. The stellar and planetary evolutionary
status is discussed in \textsection 4. Finally, our results are briefly summarized in
\textsection 5.

\section{OBSERVATIONS}

\subsection{SuperWASP Observations} 

\begin{figure}
\centering
\includegraphics[angle=0,width=0.48\textwidth,height=6.0cm]{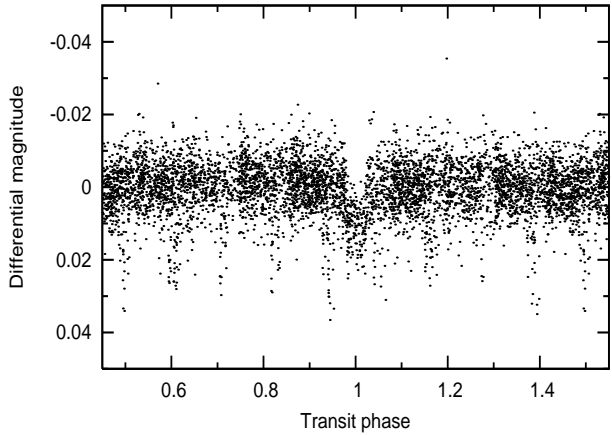}
\caption{The SuperWASP discovery light curve for the WASP-14 (1SWASPJ143306.35+215340.9). The data
points are phased using the ephemeris, $T_{\rm 0} (HJD) = 2454463.57583$, and folded on the orbital
period of 2.243752 days.}
\label{figure:wasp1}
\end{figure}

WASP-14 (= 1SWASPJ143306.35+215340.9 = USNO-B1 1118-0262485 = GSC 01482--00882) is an F5V
star with $V$ = 9.75 and $B-V$ = 0.46 in the constellation Bo\"{o}tes. It was monitored by
the SuperWASP instrument on La Palma during 2004 May to 2007 June. A detailed description
of the SuperWASP instrument, observing procedure and data-processing issues is given in
\citet{sw}, while a discussion of the candidate selection procedure is explained in Collier
Cameron et al. (2006, 2007). A total of 7338 data points were obtained for this star in
four observing runs and in two different cameras. The SuperWASP discovery light curve of
WASP-14b, as shown in Figure~\ref{figure:wasp1}, reveals a transit recurring with a period
of 2.243752 days, and a total duration of about 180 minutes between first and fourth contact. 

\subsection{Photometric follow-up}
\begin{figure}
\centering
\includegraphics[angle=0,width=0.48\textwidth,height=10.0 cm]{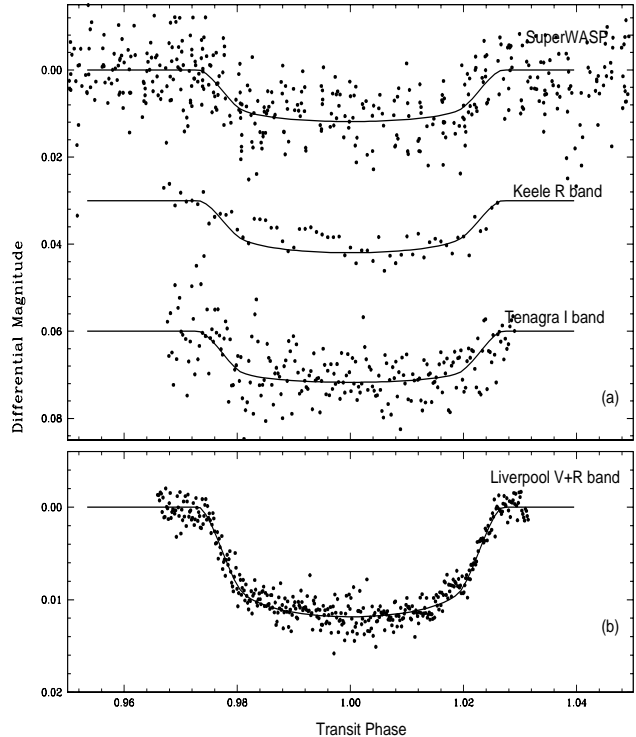}
\caption{(a) Phase folded transit light curves from the SuperWASP, Tenagra and Keele data.
(b) A high precision transit light curve of WASP-14 observed with the {\it RISE}.
The best-fitting model curves determined from a simultaneous MCMC fit
(see detail in \textsection 3.3) are shown by the solid lines.}
\label{figure:followup}
\end{figure}

Initial followup photometric observations of WASP-14 were carried out by Las Cumbres Observatory,
S. Arizona's 81-cm Tenagra robotic telescope and Keele University Observatory 60-cm telescope
in $I$ and $R$ bands respectively. All the images were bias subtracted and flat fielded before
doing aperture photometry around bright stars. Light curves along with the SuperWASP discovery
data around the transit phase are shown together with model fits in Figure~\ref{figure:followup}(a).

In order to obtain a higher precision photometry\footnote{The data is available upon request.} of
WASP-14, we observed the target field using
the high-speed imaging camera {\it RISE} mounted on the robotic 2-m Liverpool Telescope
\citep{steele,gibson}. We observed the transit on the night of 2008 May 04 in single broad band
V$+$R filter as part of the Canarian Observatories {\it International Time Programme}. The RISE
Camera consists of a thermoelectrically cooled 1024$\times$1024 pixel CCD detector which has a
pixel scale of $\sim$ 0.55 arcsec/pixel and a total field of view of $\sim$ 9.4$\times$9.4
arcmin. During this period we obtained 7200 frames of 2 sec exposures in 2$\times$2 spatial
binning. Aperture photometry was performed using the IRAF DAOPHOT routine around the target and
3 bright, non-variable comparison stars using an aperture of radius 8 pixels. We binned the
images in 30-second blocks to increase signal-to-noise ratio. To derive the differential
magnitude of WASP-14, we produced a ratio of the combined flux from the comparison stars to that
from WASP-14. We modelled the transit light curve using a Markov-chain Monte-Carlo (MCMC)
algorithm (see \textsection 3.3). The resulting normalized transit light curve of WASP-14 along
with model fit is shown in Figure~\ref{figure:followup}(b). From the model fit, we estimated a
transit depth of about 12 mmag ($(R_{\rm P}/R_{\rm *})^2 = 0.0102$) and a rms residuals of 1.4 mmag
around the best fit.

\begin{table}
\begin{center}
\caption {Radial velocities for WASP-14 from FIES/NOT and SOPHIE/OHP data. The first six data
points were taken from FIES instrument while the others were obtained with SOPHIE.}
\begin{tabular}{cccc}
\hline
Time & Rad\,Vel & $\sigma _{RV}$ & Bis\,Span  \\
(BJD-245,0000) & (km\,s$^{-1}$) & (km\,s$^{-1}$) & (km s$^{-1}$)$^{a}$  \\
\hline\\
 4461.7400 &  -5.8705  &  0.0080 &   0.0667  \\
 4462.7710 &  -4.1180  &  0.0054 &   0.0629  \\
 4465.7490 &  -4.8489  &  0.0078 &   0.0332  \\
 4466.7850 &  -5.4037  &  0.0044 &   0.0388  \\
 4490.7700 &  -5.6209  &  0.0087 &   0.0392  \\ \medskip
 4490.7820 &  -5.6383  &  0.0081 &  -0.0069  \\
 4508.5938 &  -5.3563  &  0.0112 &   0.0410  \\
 4509.5238 &  -5.0654  &  0.0114 &   0.0070  \\
 4509.5830 &  -4.8313  &  0.0098 &   0.0270  \\
 4510.5074 &  -4.5874  &  0.0129 &   0.0540  \\
 4510.5599 &  -4.7205  &  0.0106 &   0.0330  \\
 4510.6509 &  -4.8798  &  0.0099 &   0.0260  \\
 4510.6569 &  -4.8954  &  0.0107 &   0.0110  \\
 4510.6630 &  -4.9138  &  0.0111 &   0.0670  \\
 4510.6860 &  -4.9758  &  0.0100 &   0.0130  \\
 4510.7215 &  -5.0906  &  0.0108 &   0.0320  \\
 4510.7274 &  -5.1086  &  0.0112 &   0.0390  \\
 4511.5572 &  -5.6865  &  0.0107 &   0.0200  \\
 4511.6822 &  -5.3425  &  0.0114 &   0.0370  \\
 4512.5406 &  -4.2173  &  0.0109 &   0.0220  \\
 4512.5682 &  -4.2696  &  0.0125 &   0.0250  \\
 4512.6822 &  -4.4391  &  0.0107 &   0.0530  \\
 4512.7152 &  -4.5081  &  0.0103 &   0.0580  \\
 4515.6212 &  -5.8950  &  0.0099 &  -0.0070  \\
 4518.6525 &  -4.5850  &  0.0117 &   0.0340  \\
 4524.6750 &  -5.9753  &  0.0107 &   0.0700  \\
 4525.6755 &  -4.0564  &  0.0091 &   0.0560  \\
\hline
\end{tabular} \\
\label{rv}
\end{center}
$^{a}$ Bisector spans; $\sigma$$_{BS}$ $\approx$ 2.5\,$\sigma$$_{RV}$
\end{table}

\subsection{Spectroscopic follow-up} 

\begin{figure}
\centering
\includegraphics[angle=0,width=0.48\textwidth, height=11.0cm]{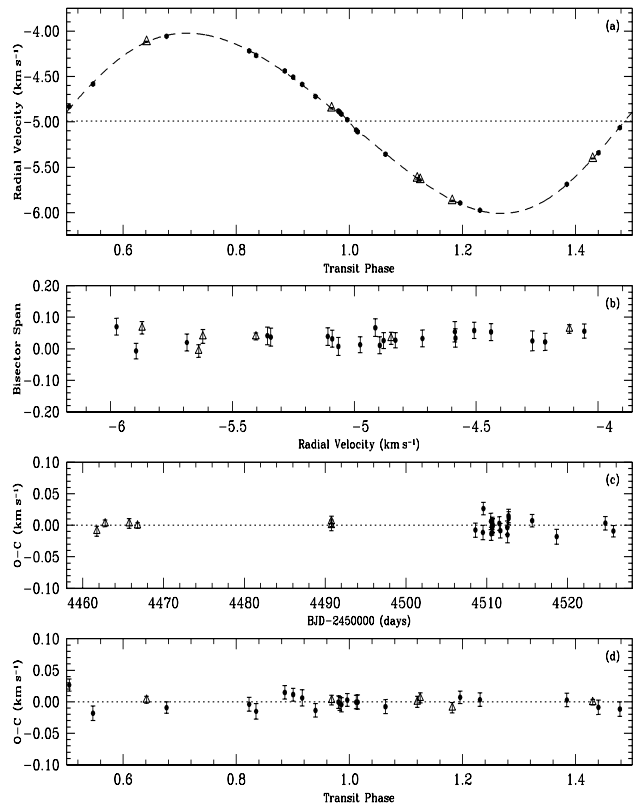}
\caption{(a) The radial-velocity curve of WASP-14. A best-fitting MCMC model
fit for the eccentric orbit is overplotted with the solid line.
(b) Bisector spans as a function of RV. Residual about the MCMC fit as a
function of transit phase (c), and as a function of time (d). The open triangle and
filled circles are observations from the FIES and SOPHIE respectively.}
\label{figure:rv}
\end{figure}

To derive the orbital parameters of the planetary system, we obtained spectroscopic observations
with the FIbre-fed Echelle Spectrograph (FIES) mounted on the 2.5-m Nordic Optical telescope.
A total of six high-resolution spectra of WASP-14 with an exposure time of 900 sec covering
the wavelength region 4000--7350 {\AA} were obtained during 2007 December 27-31 and 2008 January 25.
A further 21 RV points were secured using the SOPHIE spectrograph on the 1.93-m telescope at
observatoire de Haute-Provence during 2008 February 12-16, plus one point each on the nights
of 2008 February 19, 22, 28 and 29. A detailed description about the FIES and SOPHIE observations
is given in \citep{christian}. The journal of the FIES and SOPHIE observations is given in
Table~\ref{rv}, and includes barycentric Julian dates (BJD), RV measurements and associated error.
A phase-folded radial-velocity curve overplotted with the maximum-likelihood orbit model is shown
in the upper panel of Figure~\ref{figure:rv}(a). This shows RV  variations with a semi-amplitude
$K_1 = 0.993$~km\,s$^{-1}$ and the period derived from the photometric observations. The velocity
RMS to the orbital fit for FIES and SOPHIE are 5.8 and 11.0 m\,s$^{-1}$ respectively.

The variation of the line bisector with orbital phase is often used as a tool to detect blended
stellar companions or chromospheric activity \citep{bisector}. From the CCF we obtained the
line bisector and calculated the bisector spans as noted in the last column of the Table~\ref{rv}
and plotted, as a function of RV, in Figure~\ref{figure:rv}(b). We determined a signal-to-noise
ratio ${\rm SNR}$ = 1.45 (see, Christian et al. 2008 for detail), indicating no significant
correlation between the bisector span and RV variations. This supports the conclusion that the
motion is indeed caused by a companion planet, which given the large RV semi-amplitude, suggests
that WASP-14b is a super-massive exoplanet.

The RV residuals in Figure~\ref{figure:rv}(c) show a small dispersion of
$\sigma$(O-C) = 10.1 m\,s$^{-1}$ about the MCMC fit implying a $\chi^2$ = 23.62 for 27 degrees of
freedom. This is consistent with the error bars on radial velocity measurements. We note this is
also the order of magnitude of expected stellar `jitter' in the F5 dwarf host (see, e.g., Santos
et al.~2000, Wright~2005). We have not noticed any systematic variation in the RV residuals during
two months of our observations (see, Figure~\ref{figure:rv}(d)) hence do not suggest the presence
of any third body in the system on the basis of present data.

\subsection{Rossiter-McLaughlin effect}
Several RV observations were secured with SOPHIE during a transit of WASP-14b on the night of 2008
February 13. The relatively high impact parameter of the system breaks the degeneracy between
$v\sin i$ and the angle $\lambda$ between the projected spin axes of rotation axis of the star and
the orbit of the planet on the sky. We treated both these quantities as fitting parameters in the MCMC
code, and verified that they were indeed uncorrelated. The resulting $v\sin i = 4.7\pm1.5$ km~s$^{-1}$,
determined from the apparent radial acceleration during the transit, is consistent with
$v\sin i = 4.9\pm1.0$~km~s$^{-1}$ derived from the spectroscopic analysis, and
$v\sin i = 3.0\pm1.5$~km~s$^{-1}$ from the width of the CCF of the SOPHIE spectra which is calibrated
following the method of \citet{sant02}. We note here that FIES spectra and SOPHIE CCF provide a
$v\sin i$ averaged over all the stellar disk, whereas the R-M fit provides the $v\sin i$ at the
impact parameter of the transiting planet. The value of misalignment angle,
$\lambda = -14^{+ 21}_{-13}$ degrees, derived from the asymmetry
of the R-M effect, is consistent with zero obliquity, but is of insufficient precision to rule
out a substantial misalignment of the stellar spin and planetary orbital axes. 

\begin{figure}
\centering
\includegraphics[angle=0,width=0.48\textwidth, height=9.5cm]{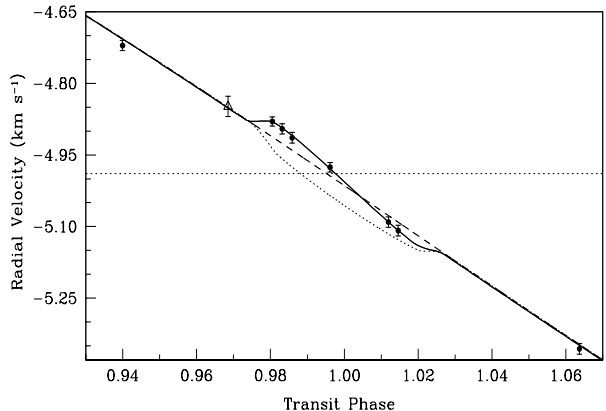}
\vspace{-4.0 cm}
\caption{The RV variations around the transit phase. Solid line shows best-fitting model fit, dashed
line shows Keplerian model without Rositter-McLaughlin effect and dotted line shows the model fit of
a transverse transit ($\lambda$ = 90$^\circ$).}
\label{figure:rm}
\end{figure}

An enlarged plot of RV variations around the transit phase is drawn in Figure~\ref{figure:rm}.
A continuous line is over-plotted for our best fit ($v\sin i = 4.7$ km~s$^{-1}$,
$\lambda = -14^{\circ}$). We also overplot Keplerian fit by the dashed line determined from
the RV measurements outside the transit in order not to be affected by the uncertainties of
the R-M effect. One can see that the first four measurements performed in the first half of
the transit are red-shifted by comparison to the Keplerian fit, whereas the last two measurements
performed in the second half of the transit are blue-shifted by comparison to the Keplerian fit.
That ordinary feature for an aligned, prograde transit shows that the R-M anomaly is indeed
significantly detected. A model of a transverse transit drawn by dotted line in the same plot
clearly exclude any variations due to $\lambda$ = 90$^\circ$. A possible transverse transit
have been detected in the case of XO-3b \citep{hebrard}, whereas HD\,147506b (HAT-P-2b) show
aligned transits \citep{winn1, loeillet}. This suggests that the history of WASP-14b is more
similar to  HAT-P-2b than the other massive planet XO-3b. Given the high eccentricity of the
system, it would be interesting to make further, more densely-sampled observations of the
R-M effect in this system. This will also allow to better constrain the system parameters,
such as $\lambda$.

\section{Determination of System Parameters}

\subsection{$T_{\rm eff}$ from the IRFM}
We used archival TYCHO, DENIS and 2MASS magnitudes to estimate the host star's effective
temperature using the Infrared Flux Method (IRFM) \citep{blackwell}. IRFM yields a nearly
model-independent determination of $T_{\rm eff}$ from the integrated stellar flux. We
individually estimated $T_{\rm eff}$ for $J$, $H$ and $K$ bands and a weighted mean was
determined for the final value of $T_{\rm eff}$. In the case of WASP-14 we find
$T_{\rm eff} = 6480\pm140$~K. It should be noted that no extinction correction was applied
in the IRFM fitting since the star is relatively nearby and out of the Galactic plane.
The $J-H = 0.22$ and $V-K = 1.13$ colours also suggest similar estimates for $T_{\rm eff}$
of $\sim$ 6230 and 6440~K respectively \citep{martin}. This suggests the star is of spectral
type of F5-7. We also determined an average angular diameter of $\theta = 0.075\pm0.004$ $mas$
from the IRFM.

\subsection{Spectral analysis}

In order to perform a detailed spectroscopic analysis of the stellar atmospheric
properties of WASP-14, we followed the same methodology as described in \citet{wasp3}.
Our results from the spectral analysis are consistent with a F5 spectral type main-sequence
star, in agreement with the F5-7 type determined from Infrared data, and the F5 type listed
in SIMBAD from the PPM catalog. In addition we also analyzed a small region around the
Li {\sc i} 6708 line to determine a lithium abundance that is defined as
$\log N({\rm Li}) \equiv 12 + ({\rm Li}/{\rm H})$ and estimated to be 2.84. The resulting
parameters from our spectral analysis are given in Table~\ref{spectral}.

\begin{table}
\centering
\caption{Stellar parameters for WASP-14. The last 6 parameters are derived from the
spectral analysis of the FIES data.}
\label{spectral}
\begin{tabular}{lll}
\hline 
R.A. (2000)               &  14$\rm^{h}$33$\rm^{m}$06.35$\rm^{s}$\\
DEC (2000)                &  +21$^\circ$53$^{'}$40.9$^{''}$	\\ 
$V$ (mag)                 & 9.75		\\
$B-V$ (mag)               & 0.46$\pm$0.03	\\
Age (Gyr)                 & $\sim$ 0.5--1.0	\\ \medskip
Distance (pc)             & 160$\pm$20  	\\ 
Spectral Type             & F5V 		\\ 
$v\sin i$  (km\,s$^{-1})$ & 4.9$\pm$1.0 	\\
$T_{\rm eff}$ (K)         & 6475$\pm$100	\\ 
log ${\it g}$ (cgs)       & 4.07$\pm$0.2	\\ 
$[$M/H$]$ (dex)           & 0.0 $\pm$0.2	\\ 
$\log N({\rm Li})$        & 2.84$\pm$0.05	\\ \hline
\end{tabular}
\end{table}

\begin{table}
\centering
\caption{Planetary and stellar parameters for the WASP-14 system derived from the MCMC analysis along with
$1\sigma$ limits.}
\label{mcmc}
\small
\begin{tabular}{lcl}
\hline\medskip
Parameter & Symbol & Values \\ \hline 
Transit epoch  (HJD)      & $T_{\rm 0}$                & 2454463.57583  $ \pm0.00053                $ ~days\\ 
Orbital Period            & $P$                        &       2.243752 $ \pm0.000010               $ ~days\\ 
Transit duration          & $T$                        &       0.1275   $ ^{+0.0028  }_{-0.0031  }  $ ~days\\ 
Transit Depth             & $(\frac{R_{\rm P}}{R_{\rm *}})^2$  & 0.0102 $ ^{+0.0002  }_{-0.0003  }  $ ~mag \\  \medskip
Impact parameter          & $b$                        &      0.535	$ ^{+0.031   }_{-0.041   }  $ ~R$_{\odot}$\\
Orbital Inclination       & $i$                        &     84.32	$ ^{+0.67    }_{-0.57	 }  $ ~deg \\ 
Orbital semi-major axis   & $a$                        &      0.036	$ \pm0.001                  $ ~AU \\ 
Orbital eccentricity      & $e$                        &      0.091	$ \pm0.003                  $	 \\ 
Arg. periastron           & $\omega$                   &   -106.629	$ ^{+0.693   }_{-0.678   }  $ ~rad \\ 
Stellar Reflex velocity   & $K_{\rm 1}$                &      0.993	$ \pm0.003                  $ ~km~s$^{-1}$ \\  \medskip
Center-of-mass velocity   & $\gamma$                   &     -4.990	$ \pm0.002                  $ ~km~s$^{-1}$\\ 
Stellar mass              & $M_{\rm *}$                &      1.211	$ ^{+0.127   }_{-0.122   }  $ ~M$_{\odot}$\\ 
Stellar radius            & $R_{\rm *}$                &      1.306	$ ^{+0.066   }_{-0.073   }  $ ~R$_{\odot}$ \\  
Stellar density           & $\rho_{\rm *}$             &      0.542	$ ^{+0.079   }_{-0.060   }  $ ~$\rho_{\odot}$\\  \medskip
Stellar surface gravity   & $\log g_{*}$               &      4.287	$ ^{+0.043   }_{-0.038   }  $ ~[cgs]\\ 
Planet mass               & $M_{\rm P}$                &      7.341	$ ^{+0.508   }_{-0.496   }  $ ~M$_{\rm J}$\\ 
Planet radius             & $R_{\rm P}$                &      1.281	$ ^{+0.075   }_{-0.082   }  $ ~R$_{\rm J}$\\ 
Planet density            & $\rho_{\rm P}$             &      3.501	$ ^{+0.636   }_{-0.495   }  $ ~$\rho_{\rm J}$\\ 
Planetary surface gravity & $\log g_{\rm P}$           &      4.010	$ ^{+0.049   }_{-0.042   }  $ ~[cgs] \\ \medskip
Planet temp ($A = 0$)     & $T_{\rm P}$                &   1866.12	$ ^{+36.74   }_{-42.09   }  $ ~K    \\ 
Photometric data points &   $N_{df}$                   &   6482                                             \\
$\chi^2$ (photometric)    & $\chi^2_{phot}$            &   6452.78                                          \\
Spectroscopic data points & $N_{df}$                   &   27                                               \\
$\chi^2$ (spectroscopic)  & $\chi^2_{spec}$            &     23.62                                          \\
 \hline
\end{tabular}
\end{table}

\subsection{Model fit to determine planetary parameters}
Transit photometry combined with radial velocity measurements provides detailed information
about the planetary orbit and the stellar and planetary parameters. High precision photometric
data was combined with the SOPHIE radial-velocity measurements in a simultaneous MCMC analysis
to find the parameters of the WASP-14 system and their covariance matrix. Our implementation
of MCMC for transiting exoplanets is presented in detail in \citet{waspmcmc}, with an extension
to fit eccentric orbits described in \citet{wasp3}. Various parameters determined from model
fit along with the corresponding 1-$\sigma$ error limits, are listed in Table~\ref{mcmc}.
Since a zero eccentricity orbit is expected for such a short-period planet system due to
relatively small tidal circularisation timescale (see \textsection 4.2), we have also analysed
the model assuming a zero-eccentricity orbit which has increased $\chi^2$ by about 370.
Using the stellar radius given in the table and angular diameter of the star determined from
the IRFM, we estimated a distance of about $160\pm20$ pc for the host star.

\section{Evolutionary Status}

\subsection{Stellar and planetary evolution}

\begin{figure}
\centering
\includegraphics[angle=0,width=0.48\textwidth,height=6.0cm]{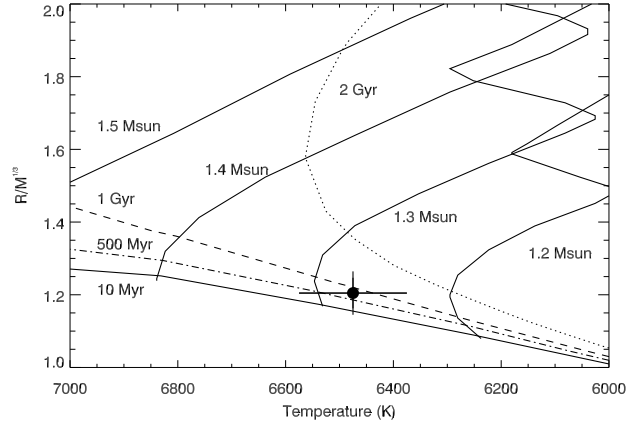}
\caption{Position of WASP-14 in $R/M^{1/3}$ -- $T_{\rm eff}$ plane. \citet{girardi}
solar metallicity stellar evolutionary tracks for different masses and isochrones of different
ages are also shown.}
\label{figure:stevo}
\end{figure}

We estimated the age of WASP-14 using a maximum-likelihood fit to the stellar evolutionary
tracks for low- and intermediate-mass stars given by \citet{girardi}. In Figure~\ref{figure:stevo},
we have shown position of WASP-14 in the $R/M^{1/3}$ versus $T_{\rm eff}$ plane. The evolutionary
tracks of different stellar mass along with the isochrones of different ages taken from
\citet{girardi} models are also drawn. This shows the star with $T_{\rm eff}$ = 6475$\pm$100~K
and a radius of 1.30$\pm$0.07~R$_\odot$, is consistent with being a main-sequence object and has
an interpolated mass of 1.21 $M_\odot$ from the models. The MCMC fit to the photometry and
radial velocities (Table~\ref{mcmc}) was computed using this value for the stellar mass, ensuring
consistency between the evolutionary status and the system parameters obtained from the MCMC fit. 
The resulting value of $\log g = 4.33\pm0.06$ is higher but more precisely determined 
than that obtained from the spectral fit, and is consistent within the uncertainty of the 
latter measurement. Although there is significant uncertainty in the effective temperature
of the star, WASP-14 seems to be consistent with being an approximate age of 0.5--1.0 Gyr. 

In F-type stars the convection zone does not penetrate deep enough to transport lithium
to regions with temperatures above 6000~K, reducing the efficiency of lithium depletion,
with the exception of the Li gap or `Boesgaard gap' \citep{boesgaard, balachandran}.
The temperature of WASP-14, $T_{\rm eff}$ = 6475$\pm$100~K, places it close to the Li gap,
where lithium depletion is enhanced by an as-yet unknown mechanism \citep{lithgap}. Although
the rate of lithium depletion in the Li gap is not very well calibrated, the relatively high
lithium content $\log N({\rm Li})$ = 2.84$\pm$0.05 and rotation speed $v\sin i = 4.9\pm1.0$
km~s$^{-1}$ possibly indicate that WASP-14 is a young star. Similar high levels of lithium
is also found in the Hyades cluster (age $\sim$ 600 Myr) with
$\log N({\rm Li})$ = 2.77$\pm$0.21 for stars with $T_{\rm eff} = 6200\pm150$~K \citep{sestito}.

We further compare our results with the models of \citet{fortney} where masses and radii
are given for a range of planetary masses, orbital separations and stellar ages.
For WASP-14b, these models predict a planetary radius $< 1.20~R_{\rm J}$
for a host star of  age  1 Gyr, while for an age of 300 Myr, the radius is predicted to be about
1.20-1.26 $R_{\rm J}$. A radius of $1.28\pm0.08 ~R_{\rm J}$ for the planet suggests
that WASP-14 possibly lies somewhere in the middle of these two age limits -- in agreement with
our earlier finding. For massive planets ($> 5~M_{\rm J}$), the model shows $<$3\% change
in planet radius with cores of mass from 0--100 $M_\oplus$. In the context of \citet{fortney}
model our observations do not place strong constraints on the core mass of WASP-14b.

\subsection{Tidal evolution}

One of the interesting features of WASP-14b is its high orbital eccentricity, $e=0.091$, for
its small orbital distance of 0.036 AU. At this distance, any orbital eccentricity should induce 
strong tidal energy dissipation within the planet leading to circularisation of the orbit.
\citet{matsumura} determined a median eccentricity of 0.013 for close-in planets with semi-major
axis $a < 0.1$ AU. The nearly-circular orbits of most close-orbiting exoplanets indicate tidal
circularisation  timescales significantly shorter than the system age. The high eccentricity
of WASP-14b may thus indicate either a system age comparable to the tidal circularisation
timescale, or the presence of an additional perturbing body in the system.

In the tidal evolution scenario originally developed by \citet{goldreich} and more recently
discussed by \citet{dobbsdixon,jackson} and others, the circularisation timescale ($\tau_e$)
and spiral-in timescale ($\tau_a$) for a single planet are given by the logarithmic time
derivatives of the orbital eccentricity and semi-major axis:
$$\tau_e = e / \dot{e};~~~~~~~ \tau_a = a / \dot{a}$$
Using the equations given in \citet{jackson} for close-in extrasolar planets,
and the planetary and stellar parameters for the WASP-14 system, we obtain
$\tau_e \simeq 1 / [28.5/(Q_p/10^5) + 6.6/(Q_s/10^6)] $ Gyr and
$\tau_a \simeq 1 / [0.5165(e/.089)^2/(Q_p/10^5) + 2.359/(Q_s/10^6)] $ Gyr.
Here the tidal dissipation parameters $Q_p$ for the planet and $Q_s$ for the star are assigned
fiducial values of $10^5$ and $10^6$ respectively in accordance with the best-fitting values
derived from the study of \citet{jackson}. It should be mentioned here that there could be
considerable uncertainties in both these parameters; moreover, the timescales are inversely
proportional to the fifth powers of the planetary and stellar radii, both of which contain
significant uncertainties in their determinations.

From the expression for $\tau_e$ above, the circularisation timescale of the planet's orbit
will be less than 1 Gyr if either $Q_p < 3.7\times 10^6$ or $Q_s < 6.6\times 10^6$. The
timescale for decay of the orbital semi-major axis of the planet via torques arising from the
tidal bulge raised on the star could also be less than 1 Gyr if $Q_s < 2.4\times 10^6$. These
same torques might be expected to spin the star up on a timescale of the order of 1 Gyr, since
the angular momentum loss in the wind of a F star like WASP-14 is quite weak. A dissipation
factor $Qs > 6.6\times 10^6$ could increase the life expectancy of the planet to a value
similar to the main-sequence lifetime of the F-type host.

If $Q_p$ is low and the circularisation timescale is short, the most plausible alternative
mechanism for maintaining the high eccentricity of WASP-14b is secular interaction with an
additional planet in the system (see, e.g., Adams \& Laughlin~2006). From the arguments above,
however, it is not clear whether an additional planet is needed to maintain the observed
eccentricity over the $\sim$ 1 Gyr lifetime of the star. This is only needed if the rate at
which tidal energy is dissipated in the planet is high, with $Q_p < 3 \times 10^6$. The slow
stellar spin suggests that magnetic braking is dominating over torques arising from the tidal
bulge raised on the star. This in turn suggests a low rate of dissipation in the star, with
$Q_s > 10^8$ or so. A further discussion of eccentric orbits in close-in exoplanets and their
constraints on Q can be found in \citet{matsumura}.

Two other recently announced planets WASP-10b (e=0.059; Christian et al. 2008) and WASP-12b
(e=0.064; Hebb et al. 2008) also have apparently significant orbital eccentricity. While
\citet{shen} noted that sparsely sampled radial velocity curves are often interpreted
as having significant orbital eccentricity, in the case of the WASP planets the spectroscopic
data is usually obtained over a small number of cycles and with good precision. Consequently,
we expect the eccentricities derived for these systems to be free from bias or selection effects.

\section{Discussion}
We have discovered a new massive exoplanet WASP-14b orbiting with a period of about 2.244 days
around its host main-sequence star GSC 01482--00882. Spectral analysis of the star implies a
spectral type of F5V with solar metallicity. High precision photometric and spectroscopic
follow-up observations reveal that the planet has a mass of 7.3$\pm$0.5 M$_{\rm J}$,
radius of 1.28$\pm$0.08 R$_{\rm J}$ suggesting a mean density of 
3.5$\pm$0.6 $\rho_{\rm J}$ ($\sim$ 4.6 g\,cm$^{-3}$), and an eccentricity of
0.091$\pm$0.003. The mean density of WASP-14b is high in comparison with a typical
Hot Jupiter density of 0.34-1.34 g\,cm$^{-3}$ \citep{loeillet}, and similar to that of rocky
planets and makes it densest transiting exoplanets so far discovered with $<$ 3 d orbital period.
The planet is too massive to fit the mass-period relation given by \citet{torres}, but its
radius is consistent with the theoretical radius expected from the \citet{fortney} model.
Spectral analysis reveals that WASP-14 has a high lithium abundance, $\log N({\rm Li})$ = 2.84.
This is consistent with its stellar temperature ($T_{\rm eff}$ = 6475$\pm$100 K) and age of
about 0.5--1.0 Gyr which is further supported by the fitting of stellar evolutionary models
of the \citet{girardi}.

High orbital eccentricities in close-in planets are often explained by perturbations from
a third body (Jackson et al. 2008 and references therein) or internal structures that result
in their tidal dissipation factor being significantly larger that that commonly assumed
$Q_p \sim 10^6$ (Matsumura et al. 2008, Hebb et al. 2008). Although we do not see any systematic
variation in RV residuals during our two months of spectroscopic observations of WASP-14, we
expect that long term radial velocity monitoring will help constrain the nature of the third
body in this systems hence make it an important target both for future transit-timing variation
studies and for longer-term RV monitoring to establish the mass and period of the putative
outer planet.

WASP-14b is one of the most massive transiting planets known along with HAT-P-2b
\citep{hat2,winn1, loeillet} and XO-3b \citep{christopher,winn2,hebrard} and its physical
characteristics closely resemble with those of HAT-P-2 except the latter has a much longer
orbital period and smaller radius. However, there is still no firm explanation about the
formation of such highly massive planets (e.g., H\'{e}brard et. al~2008). A quite interesting
feature is that all these three massive exoplanets has unusually large eccentric orbit for
their short orbital period. Matsumura et al. (2008) argues that the new class of eccentric,
short period transiting planets are still in the process of circularization and speculates
that $Q_p$ could be as large as 10$^9$ for these planets. 

The success of theoretical models of planetary structure depends heavily on our precise knowledge
of the basic physical parameters of the planetary systems. Transit surveys in recent times have
produced a large sample of transiting planets that show a remarkable diversity in their mass,
radius and internal structure. While most of the Jupiter-mass transiting planets are well
explained by existing models, planets which show excessive mass for their small radius or
relatively low mass for their large radius are yet to be explained satisfactorily by any available
model.  WASP-14b is one of the few massive exoplanets, some of which are even in closer orbits
than most other hot Jupiters. It poses a great challenge for theoretical models to explain their
internal structure, atmospheric dynamics and heat distribution.

\section*{Acknowledgments}
The SuperWASP Consortium consists of astronomers primarily from the Queen's University Belfast,
St Andrews, Keele, Leicester, The Open University, Isaac Newton Group La Palma and
Instituto de  Astrof{\'i}sica de Canarias. SuperWASP Cameras were constructed
and operated with funds made available from Consortium Universities and the UK's Science and
Technology Facilities Council. SOPHIE observations have been funded by the Optical Infrared
Coordination Network. Data from the Liverpool and NOT telescopes was obtained under the
auspices of the International Time of the Canary Islands. We extend our thanks to the staff
of the ING and OHP for their continued support of SuperWASP-N and SOPHIE instruments. FPK is
grateful to AWE Aldermaston for the award of a William Penney Fellowship.

\label{lastpage}
\end{document}